\documentclass[
    reprint,
    superscriptaddress,
    aps, prl,
    amsmath,amssymb,
    twocolumn,
    showkeys,
    floatfix,
    nofootinbib,
]{revtex4-2}

\usepackage{graphicx}
\usepackage{xspace}
\usepackage{isotope}
\usepackage{orcidlink} 
\usepackage{lipsum}
\usepackage{siunitx}
\usepackage{upgreek}
\usepackage{booktabs}  
\usepackage{float}
\usepackage{colortbl}
\usepackage{amsmath}
\usepackage{array}
\usepackage[mathlines]{lineno} 
\usepackage[capitalise]{cleveref}
\crefname{figure}{Figure}{Figures}
\crefname{table}{Table}{Tables}

\begin{document}



\title{WIMP Dark Matter Search using a 3.1 Tonne-Year Exposure of the XENONnT Experiment}
\newif\ifshowtldr
\showtldrfalse

\newcommand*{\comment}{\textcolor{red}}
\newcommand*{\needs}{\textcolor{purple}}
\newcommand*{\todo}{\textcolor{brown}}
\newcommand*{\software}{\textsc}
\newcommand*{\tldr}[1]{
    \ifshowtldr
        \textit{\textcolor{gray}{tldr: #1..}}
    \fi
}

\newcommand{\x}{\ensuremath{\mathrm{X}}\xspace}
\newcommand{\y}{\ensuremath{\mathrm{Y}}\xspace}
\newcommand{\xy}{\ensuremath{\mathrm{XY}}\xspace}
\newcommand{\z}{\ensuremath{\mathrm{Z}}\xspace}
\newcommand{\R}{\ensuremath{\mathrm{R}}\xspace}
\newcommand{\radius}{\ensuremath{\mathrm{R}}\xspace}

\newcommand{\sone}{\ensuremath{\mathrm{S1}}\xspace}
\newcommand{\stwo}{\ensuremath{\mathrm{S2}}\xspace}
\newcommand{\csone}{\ensuremath{\mathrm{cS1}}\xspace}
\newcommand{\cstwo}{\ensuremath{\mathrm{cS2}}\xspace}
\newcommand{\srzero}{\ensuremath{\mathrm{SR0}}\xspace}
\newcommand{\srone}{\ensuremath{\mathrm{SR1}}\xspace}
\newcommand{\sronea}{\ensuremath{\mathrm{SR1a}}\xspace}
\newcommand{\sroneb}{\ensuremath{\mathrm{SR1b}}\xspace}

\newcommand{\gone}{\ensuremath{g_1}\xspace}
\newcommand{\gtwo}{\ensuremath{g_2}\xspace}

\newcommand{\lightyield}{\ensuremath{\mathrm{LY}}\xspace}
\newcommand{\chargeyield}{\ensuremath{\mathrm{CY}}\xspace}

\newcommand{\rntwotwozero}{\ensuremath{^{220}\mathrm{Rn}}\xspace}
\newcommand{\xeonethreesix}{\ensuremath{^{136}\mathrm{Xe}}\xspace}
\newcommand{\xeonetwofour}{\ensuremath{^{124}\mathrm{Xe}}\xspace}
\newcommand{\arthreeseven}{\ensuremath{^{37}\mathrm{Ar}}\xspace}

\newcommand{\evcc}{\ensuremath{\mathrm{eV}/c^2}\xspace}
\newcommand{\kevcc}{\ensuremath{\mathrm{keV}/c^2}\xspace}
\newcommand{\mevcc}{\ensuremath{\mathrm{MeV}/c^2}\xspace}
\newcommand{\gevcc}{\ensuremath{\mathrm{GeV}/c^2}\xspace}
\newcommand{\tevcc}{\ensuremath{\mathrm{TeV}/c^2}\xspace}
\newcommand{\kever}{\ensuremath{\mathrm{keV_{ER}}}\xspace}
\newcommand{\kevnr}{\ensuremath{\mathrm{keV_{NR}}}\xspace}

\newcommand{\wimp}{\ensuremath{\mathrm{WIMP}}\xspace}
\newcommand{\cevns}{CE\ensuremath{\nu}NS\xspace}
\newcommand{\dec}{\ensuremath{2\upnu\mathrm{ECEC}}\xspace}
\newcommand{\toymc}{toy-MC\xspace}
\newcommand{\toymcs}{toy-MCs\xspace}

\newcommand{\bologna}{\affiliation{Department of Physics and Astronomy, University of Bologna and INFN-Bologna, 40126 Bologna, Italy}}
\newcommand{\chicago}{\affiliation{Department of Physics, Enrico Fermi Institute \& Kavli Institute for Cosmological Physics, University of Chicago, Chicago, IL 60637, USA}}
\newcommand{\coimbra}{\affiliation{LIBPhys, Department of Physics, University of Coimbra, 3004-516 Coimbra, Portugal}}
\newcommand{\columbia}{\affiliation{Physics Department, Columbia University, New York, NY 10027, USA}}
\newcommand{\lngs}{\affiliation{INFN-Laboratori Nazionali del Gran Sasso and Gran Sasso Science Institute, 67100 L'Aquila, Italy}}
\newcommand{\mainz}{\affiliation{Institut f\"ur Physik \& Exzellenzcluster PRISMA$^{+}$, Johannes Gutenberg-Universit\"at Mainz, 55099 Mainz, Germany}}
\newcommand{\mpik}{\affiliation{Max-Planck-Institut f\"ur Kernphysik, 69117 Heidelberg, Germany}}
\newcommand{\munster}{\affiliation{Institut f\"ur Kernphysik, University of M\"unster, 48149 M\"unster, Germany}}
\newcommand{\nikhef}{\affiliation{Nikhef and the University of Amsterdam, Science Park, 1098XG Amsterdam, Netherlands}}
\newcommand{\nyuad}{\affiliation{New York University Abu Dhabi - Center for Astro, Particle and Planetary Physics, Abu Dhabi, United Arab Emirates}}
\newcommand{\purdue}{\affiliation{Department of Physics and Astronomy, Purdue University, West Lafayette, IN 47907, USA}}
\newcommand{\rice}{\affiliation{Department of Physics and Astronomy, Rice University, Houston, TX 77005, USA}}
\newcommand{\stockholm}{\affiliation{Oskar Klein Centre, Department of Physics, Stockholm University, AlbaNova, Stockholm SE-10691, Sweden}}
\newcommand{\subatech}{\affiliation{SUBATECH, IMT Atlantique, CNRS/IN2P3, Nantes Universit\'e, Nantes 44307, France}}
\newcommand{\torino}{\affiliation{INAF-Astrophysical Observatory of Torino, Department of Physics, University  of  Torino and  INFN-Torino,  10125  Torino,  Italy}}
\newcommand{\ucsd}{\affiliation{Department of Physics, University of California San Diego, La Jolla, CA 92093, USA}}
\newcommand{\wis}{\affiliation{Department of Particle Physics and Astrophysics, Weizmann Institute of Science, Rehovot 7610001, Israel}}
\newcommand{\zurich}{\affiliation{Physik-Institut, University of Z\"urich, 8057  Z\"urich, Switzerland}}
\newcommand{\paris}{\affiliation{LPNHE, Sorbonne Universit\'{e}, CNRS/IN2P3, 75005 Paris, France}}
\newcommand{\freiburg}{\affiliation{Physikalisches Institut, Universit\"at Freiburg, 79104 Freiburg, Germany}}
\newcommand{\napels}{\affiliation{Department of Physics ``Ettore Pancini'', University of Napoli and INFN-Napoli, 80126 Napoli, Italy}}
\newcommand{\nagoya}{\affiliation{Kobayashi-Maskawa Institute for the Origin of Particles and the Universe, and Institute for Space-Earth Environmental Research, Nagoya University, Furo-cho, Chikusa-ku, Nagoya, Aichi 464-8602, Japan}}
\newcommand{\laquila}{\affiliation{Department of Physics and Chemistry, University of L'Aquila, 67100 L'Aquila, Italy}}
\newcommand{\tokyo}{\affiliation{Kamioka Observatory, Institute for Cosmic Ray Research, and Kavli Institute for the Physics and Mathematics of the Universe (WPI), University of Tokyo, Higashi-Mozumi, Kamioka, Hida, Gifu 506-1205, Japan}}
\newcommand{\kobe}{\affiliation{Department of Physics, Kobe University, Kobe, Hyogo 657-8501, Japan}}
\newcommand{\kit}{\affiliation{Institute for Astroparticle Physics, Karlsruhe Institute of Technology, 76021 Karlsruhe, Germany}}
\newcommand{\tsinghua}{\affiliation{Department of Physics \& Center for High Energy Physics, Tsinghua University, Beijing 100084, P.R. China}}
\newcommand{\ferrara}{\affiliation{INFN-Ferrara and Dip. di Fisica e Scienze della Terra, Universit\`a di Ferrara, 44122 Ferrara, Italy}}
\newcommand{\groningen}{\affiliation{Nikhef and the University of Groningen, Van Swinderen Institute, 9747AG Groningen, Netherlands}}
\newcommand{\westlake}{\affiliation{Department of Physics, School of Science, Westlake University, Hangzhou 310030, P.R. China}}
\newcommand{\shenzhen}{\affiliation{School of Science and Engineering, The Chinese University of Hong Kong (Shenzhen), Shenzhen, Guangdong, 518172, P.R. China}}
\newcommand{\coimbrapoli}{\affiliation{Coimbra Polytechnic - ISEC, 3030-199 Coimbra, Portugal}}
\newcommand{\uniheidelberg}{\affiliation{Physikalisches Institut, Universit\"at Heidelberg, Heidelberg, Germany}}
\newcommand{\roma}{\affiliation{INFN-Roma Tre, 00146 Roma, Italy}}
\newcommand{\bucknell}{\affiliation{Department of Physics \& Astronomy, Bucknell University, Lewisburg, PA, USA}}

\author{E.~Aprile\,\orcidlink{0000-0001-6595-7098}}\columbia
\author{J.~Aalbers\,\orcidlink{0000-0003-0030-0030}}\groningen
\author{K.~Abe\,\orcidlink{0009-0000-9620-788X}}\tokyo
\author{S.~Ahmed Maouloud\,\orcidlink{0000-0002-0844-4576}}\paris
\author{L.~Althueser\,\orcidlink{0000-0002-5468-4298}}\munster
\author{B.~Andrieu\,\orcidlink{0009-0002-6485-4163}}\paris
\author{E.~Angelino\,\orcidlink{0000-0002-6695-4355}}\torino\lngs
\author{D.~Ant\'on~Martin\,\orcidlink{0000-0001-7725-5552}}\chicago
\author{S.~R.~Armbruster\,\orcidlink{0009-0009-6440-1210}}\mpik
\author{F.~Arneodo\,\orcidlink{0000-0002-1061-0510}}\nyuad
% ...
\collaboration{XENON Collaboration}\email[]{xenon@lngs.infn.it}\noaffiliation
\date{\today}

\begin{abstract}
We report on a search for weakly interacting massive particle (WIMP) dark matter (DM) via elastic DM-xenon-nucleus interactions in the XENONnT experiment.
We combine datasets from the first and second science campaigns resulting in a total exposure of 3.1 tonne-years.
In a blind analysis of nuclear recoil events with energies above $3.8\,\kevnr$, we find no significant excess above background.
We set new upper limits on the spin-independent WIMP-nucleon scattering cross section for WIMP masses above 10\,\gevcc with a minimum of $1.7\,\times\,10^{-47}\,\mathrm{cm^2}$  at 90\,\% confidence level for a WIMP mass of $30$\,\gevcc.
We achieve a best median sensitivity of $1.4\,\times\,10^{-47}\,\mathrm{cm^2}$ for a $41$\,\gevcc WIMP.
Compared to the result from the first XENONnT science dataset, we improve our sensitivity by a factor of up to 1.8.
\end{abstract}
\maketitle

\textit{Introduction\textemdash}Observational evidence from galactic to cosmic scales indicates the existence of massive, non-baryonic dark matter (DM) in the Universe~\cite{Bertone:2004pz}.
Among numerous models of DM, weakly interacting massive particles (WIMPs) in the mass range between \gevcc and a few \tevcc are one of the most promising and physics-motivated DM candidates, which are naturally predicted by several extensions of the standard model~\cite{Roszkowski:2017nbc}.
Dual-phase liquid xenon (LXe) time projection chambers (TPCs) are currently the most sensitive experiments directly searching for these particles.
They have placed stringent upper limits on cross sections for elastic spin-independent (SI)  WIMP-nucleon interactions~\cite{XENON:2023wimp, LZ:2025_wimp, PandaX:2025_wimp_search}.

The XENONnT~\cite{XENON:2024instr} experiment is operated underground at the INFN Laboratori Nazionali del Gran Sasso (LNGS).
The experiment consists of three nested detectors: the central LXe TPC housed in a cryostat is enclosed by a neutron veto (NV) detector~\cite{XENON:2024neutron_veto} which is situated within, but optically separated from, a muon veto detector~\cite{XENON:2014mveto}.
Both veto detectors are inside a 700-t water tank and function as water Cherenkov detectors.
All data used in this Letter were acquired with demineralized water, relying on neutron capture on hydrogen as in \cite{XENON:2023wimp}. 

The cylindrical TPC is immersed in \SI{8.5}{\tonne} of LXe, with gaseous xenon (GXe) on top.
Particle interactions in the LXe lead to prompt scintillation light as well as ionization electrons.
The light is detected by arrays of photomultiplier tubes (PMTs) at both ends of the cylinder.
Electrons are moved toward the liquid surface by an electric drift field, where a stronger extraction field accelerates the electrons into the GXe, leading to a drift-delayed proportional scintillation signal.
The measured prompt and delayed light signals are denoted S1 and S2, respectively.
The S1--S2 combination allows for energy and three-dimensional position reconstruction.
Compared to electronic recoils (ERs), mainly expected from backgrounds, nuclear recoils (NRs) from WIMP scattering feature smaller S2/S1 ratios, allowing for ER/NR discrimination~\cite{Aprile:2006kx}.

The sensitive volume of the detector has a diameter of \SI{1.33}{m}, a maximum electron drift length of \SI{1.49}{m}, and contains \SI{5.9}{\tonne} of LXe.
All detector construction materials were selected for low radioactivity~\cite{XENON:2021radiopurity}.
The walls are made of polytetrafluoroethylene (PTFE) and cover the inside of the electric field cage~\cite{XENONnT:2023field}.
A drift field of \SI{23}{\volt/\cm} is established between a cathode electrode at the bottom of the active volume and a gate electrode just below the LXe surface.
The extraction field is set between the gate and the anode electrode in the gas phase ($\SI{2.9}{\kilo\volt/\cm}$ in the liquid).
All electrodes are composed of parallel wires.
Two (four) transverse wires support the gate (anode) wires against sagging.
Additional parallel-wire screening electrodes protect the PMT arrays, which contain 494 Hamamatsu R11410-21 $3\,\mathrm{in.}$ PMTs~\cite{Antochi:2021pmt}.
PMT pulses above predefined digitization thresholds are recorded with a triggerless data acquisition system~\cite{XENON:2023daq}, stored, and further processed using the software \software{strax(en)} \cite{strax,straxen}.
Electronegative impurities that affect electron drift, and radon emanating from surfaces are continuously removed from xenon via gas+liquid purification and online distillation, respectively~\cite{XENON:2024liquidpurification, Plante:2022khm, Murra:2022mlr, XENON:2024instr}.
\isotope[85]{Kr} was removed via cryogenic distillation as well at the start of the experiment.

\tldr{livetime, pressure, temperature, liquid level etc.}
\textit{Dataset\textemdash}In this Letter, 95.1 days of data from the first science run (SR0) of XENONnT, already published in \cite{XENON:2023wimp}, were combined with new data from the second science run (SR1) that lasted from May 19$^{\text{th}}$, 2022 to August 8$^{\text{th}}$, 2023.
The WIMP signal region of the SR1 data was blinded as in \cite{XENON:2023wimp} until the full analysis procedure had been fixed, while the SR0 data of the previous blind analysis were kept untouched.

During the preparation for SR1, a small amount of xenon with commercial-grade purity was accidentally injected into the system without prior distillation, resulting in increased ER background levels from $^{85}\mathrm{Kr}$ and $^{37}\mathrm{Ar}$.
Rare gas mass spectrometry~\cite{Lindemann:2013kna} of xenon samples indicated a molar concentration of a few parts per trillion $^\text{nat}\text{Kr}$/Xe, which is about a factor 60 higher than the usual level.
This initial period of SR1 with an elevated ER background rate is called \sronea and includes one month of cryogenic distillation that reduced the ER background level.
The subsequent low-background period is referred to as \sroneb.
The total live time for SR1 is $186.5\;\text{days}$ ($66.6+119.9\;\text{days}$ for \sronea+\xspace\sroneb).
Temperature, pressure, and liquid level remained stable at $(177.2 \pm 0.4)\,\mathrm{K}$, $(1.92 \pm 0.02)\,\mathrm{bar}$ and $(4.8 \pm 0.2)\,\mathrm{mm}$.
The liquid level in SR1 was lowered compared to SR0 by $\SI{0.2}{\mm}$, which mitigated the occurrence of localized bursts of single electron (SE) emission from the top electrodes at high rates, referred to as hot spot.
With a $\SI{50}{\V}$ anode voltage increase, the resulting SE gain of $(29.4\pm0.6)$\,PE/e$^-$ (PE denotes photoelectron) was slightly lower than the one in \srzero of 31.2\,PE/e$^-$.
No changes were made to the drift field.
The average ``electron lifetime'' (defined as the mean time for a drifting electron before being attached to an impurity) in \srone was $21.8_{-9.7}^{+6.7}\,\mathrm{ms}$.
The PMT performance was monitored with regular LED calibrations, and three additional PMTs (20 in total) were excluded from the SR1 data analysis.
In contrast to SR0, the radon distillation system was operated at its full capability in a high-flow LXe+GXe combined mode, which led to an average $^{222}\text{Rn}$ activity concentration of $\left(0.99\pm 0.01_{\text{stat}}\pm 0.07_{\text{sys}}\right)\;\upmu\text{Bq}/\text{kg}$ in \sronea and $\left(1.10\pm 0.01_{\text{stat}}\pm 0.09_{\text{sys}}\right)\;\upmu\text{Bq}/\text{kg}$ in \sroneb [with a minimum of $(0.90 \pm 0.01_\text{stat} \pm 0.07_\text{sys})\;\upmu\text{Bq}/\text{kg}$ reported in~\cite{XENON:2025radonremoval}], reducing the associated background from \isotope[214]{Pb} ground state $\upbeta$ decays by a factor of 1.9 (1.7) in \sronea (\sroneb) compared to SR0.

\tldr{XYZ reconstruction and correction.} 
An ``event'' from a particle interaction is defined by an \sone--\stwo signal pair.
The depth (\z) is reconstructed as the product of electron drift velocity and \sone--\stwo time difference. The horizontal (\x, \y) position is reconstructed from the \stwo light distribution on the top PMTs using neural network models~\cite{XENON:2024ap1}. 
The reconstructed (\x, \y, \z) position of the main signal pair is used for signal corrections. The signal reconstruction and corrections in \srone generally followed the procedures established for SR0~\cite{XENON:2024ap1}.
Since \srzero data are unchanged, we focus on the analysis changes in \srone.

The event position reconstruction in \srone was improved with an (\radius, \z)-dependent electron drift velocity from electric field simulations~\cite{XENONnT:2023field}. 
Additionally, a small charge-insensitive volume located at the outer bottom of the TPC, where electric field lines terminate on the sides of the detector, was included in the correction of the reconstructed (\x, \y, \z) positions using uniformly distributed \isotope[83\text{m}]{Kr} events in conjunction with electric field simulations~\cite{XENONnT:2023field}.

\tldr{S1 S2 reconstruction and correction, photoionization}
In \srone, we observed a higher rate of small S2 signals following large signals, attributed to photoionization on impurities in LXe~\cite{XENON:2021single_electrons}.
The delayed electron signals appeared mainly within one full TPC drift time, exhibited a correlation with the preceding signal size, and are time dependent.
The light absorption on impurities induced a percent-level time dependence in the absolute \sone signal size, as well as in the \stwo signal fraction observed by the PMTs in the bottom array.
The time dependence of both observables with respect to their averages over the whole dataset was corrected.
We believe this phenomenology is caused by non-electronegative impurities introduced by the change of mode of the radon distillation system, since no correlated change in electron lifetime was observed.
In addition, the stability of the detector response was constantly monitored by $\upalpha$ decays from \isotope[222]{Rn}, and the residual time variation was accounted for in the systematic uncertainties of the \sone and \stwo light collection efficiencies.

\begin{figure}[t]
    \centering
    \includegraphics[width=\columnwidth]{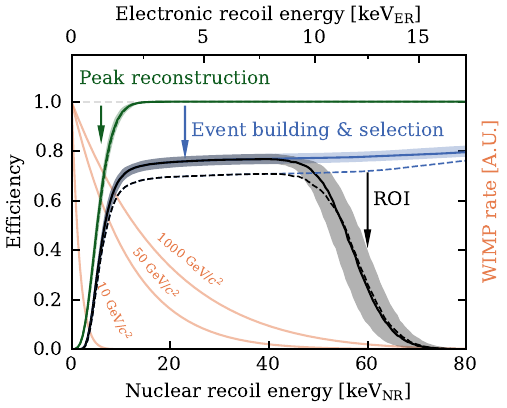}
    \caption{Efficiencies as functions of NR energy for \sronea (dashed) and \sroneb (solid). The peak reconstruction efficiency (green) is determined by the S1 threefold tight coincidence requirement. Adding event building, data selection (blue), and ROI selection (black) lowers the total efficiency. The energy range where it exceeds $10\,\%$ is $[0.6, 13.5]\,\kever$ and $[3.8, 64.1]\,\kevnr$. Recoil energy spectra for three WIMP masses without efficiencies applied are shown in orange. The upper axis shows the conversion of NR energies to the median ER energies.}
    \label{fig:sr1_nr_eff}
\end{figure}

\tldr{Energy scale}
With the corrected \sone and \stwo sizes, denoted as \csone and \cstwo, the ER energy in \kever is reconstructed as $E=W(\csone/\gone + \cstwo/\gtwo)$, with the mean energy needed to produce one observable quantum (photon or electron) of $W=13.7\,\mathrm{eV}$~\cite{Dahl_w_value}.
The photon (electron) gains $g_1$ ($g_2$), defined as the average number of detected photoelectrons per produced photon (electron), were $\gone=(0.1367\pm0.0010)\,\mathrm{PE/ph}$ and $\gtwo=(16.9\pm0.5)\,\mathrm{PE/e^{-}}$.

\tldr{Data selection}
Data selection criteria were not changed for SR0 and optimized for SR1.
A gradient-boosted decision tree (GBDT) was used to reject accidental coincidence (AC) background from incorrectly paired \sone and \stwo signals~\cite{XENON:2024ap2}.
This selection was only applied in the ``far-wire'' region ($\geq4.45\,\mathrm{cm}$ from the transverse wires, its complement is termed ``near-wire'' region), in which the \stwo pulse shape is reliably modeled due to lower distortion of the electric field~\cite{XENON:2024ap1}.
An \stwo threshold of 320\,PE was chosen to reduce the risk of AC background mismodeling, higher than the 200\,PE in \srzero due to increased AC rates caused by photoionization in \srone.
The fiducial volume (FV) retains the same shape as in SR0 but with a tighter radius cut at $58.8\,\mathrm{cm}$, enclosing ($4.00\,\pm\,0.15$)\,t of LXe.
With respect to the \srzero FV, this tighter radius further excludes 93\,\% of surface background events originating on the detector walls, while retaining 93\,\% of the WIMP signal and leaving the sensitivity nearly unchanged.

\tldr{Efficiency.}
The total efficiency consists of signal peak reconstruction, event building, event selection, and region of interest (ROI) selection efficiencies, as shown in \cref{fig:sr1_nr_eff}.
The drop in the peak reconstruction efficiency at low energies arises primarily from the threefold tight coincidence requirement for \sone signals.
It is determined via a data-driven approach and validated with Monte Carlo (MC) simulations using \software{WFSim} and \software{fuse} \cite{wfsim,fuse}.
The event-building efficiency reflects whether an event will be successfully reconstructed or obscured by, e.g., ambient SE peaks following large S2's, and depends on \sone and \stwo signal sizes.
It is determined by injecting simulated events, preselected to pass peak reconstruction, at random times into the data and processing them through the analysis pipeline using \software{saltax}~\cite{saltax} and \software{axidence}~\cite{axidence}.
The chance for an event passing the event-building process is anticorrelated with its rejection by selection criteria targeting AC events.
Therefore, the efficiency of both processes is evaluated jointly.
The event-building efficiency is lower in \sronea than in \sroneb due to a higher rate of the hot spot, and is evaluated separately for the near- and far-wire regions.
Finally, the regions of interest of both SR0 and SR1 are defined as $\csone\in[0, 100]\,\mathrm{PE}$ and $\cstwo\in[10^{2.1}, 10^{4.1}]\,\mathrm{PE}$.
The ROI efficiency uncertainty is primarily determined by the uncertainties in the fitted NR light yield (\lightyield) and charge yield (\chargeyield).
The total efficiency plateaus at $\sim71\,\%$ (77\,\%) for \sronea (\sroneb).

\textit{Signal and background models\textemdash}This analysis accounts for backgrounds from ER, NR, AC, and surface events.
An internal \isotope[220]{Rn} source (external \isotope[241]{AmBe} neutron source) is used to constrain the \lightyield and \chargeyield of ER (NR).
The ER and NR response models are parametrized and fit to the calibration datasets using a Bayesian approach \cite{XENON:2024ap2} with the software \software{appletree} \cite{appletree}, which implements an affine invariant Markov chain Monte Carlo algorithm \cite{emcee}. 
For the ER calibration, approximately 4700 events from the $\upbeta$ decay of the Rn daughter \isotope[212]{Pb} remained in the ROI after all data selections.
In SR0, \isotope[37]{Ar} ER calibration was available, enabling a better determination of $g_1$ and $g_2$ in the low-energy region, whereas in SR1 it was not.
Consequently, a combined ER fit could fail to account for the crucial uncertainty in the ER distribution in cS1--cS2 space, as \toymc studies have shown that a 1\,\% shift in the ER event distribution along \cstwo can lead to a 10\,\% change in sensitivity.
To properly capture this uncertainty, we fit ER data separately for SR0 and SR1.
For the NR calibration, a clean neutron event sample in the TPC was selected by using the NV to detect the 4.44\,MeV $\upgamma$-ray emitted from the AmBe source in coincidence with the neutron emission with a $\sim50$\,\% probability~\cite{danielwphd}.
This resulted in approximately 5700 neutron events within the ROI.
We performed a combined NR fit to both SR0 and SR1 AmBe neutron calibration datasets with shared \lightyield and \chargeyield parameters, with an updated parametrization following the NEST v2 model~\cite{Szydagis:2025}.
This allows for a better constraint on the underlying single-scatter NR response from multi-site neutron events, due to the highly spatially localized AmBe events and different source positions in SR0 and SR1.
We performed two-dimensional Poisson $\chi^2$ goodness-of-fit (GOF) tests using an equiprobable binning scheme in \csone--\cstwo space on each ER and NR best-fit model, which showed no indication of a mismatch between models and data, with the exception of the SR1 NR model with a p-value slightly below the predefined threshold.
The impact on the sensitivity of a potential mismodeling in the NR response was found to be small.

The NR response model uncertainties are parametrized as a relative WIMP signal rate uncertainty in the statistical inference.
For ER, the number of response model parameters is reduced while retaining realistic model uncertainties to make the WIMP search likelihood computationally tractable.
We use two parameters to represent the ER distribution uncertainty in \csone--\cstwo space: one from the principal component decomposition \cite{XENON:2024ap2} and another from a linear combination of \gone and \gtwo with a correlation coefficient.
These two shape parameters are propagated to the statistical inference of the results.

\tldr{WIMP signal..}
For the SI WIMP model, the energy spectrum is based on the Helm form factor~\cite{helm1956inelastic} and the standard halo model parameters as suggested in~\cite{Baxter:2021pqo}.

\tldr{ER background..}
The dominant background in this analysis is from ER interactions.
The contribution from $\upbeta$ decays of \isotope[214]{Pb}, which constituted the primary ER background in SR0, together with $\upbeta$ decays from \isotope[85]{Kr}, $\upgamma$-ray background from detector materials, and solar neutrino-electron scattering, exhibits an approximately flat energy spectrum within the ROI.
The double $\upbeta$ decays of \isotope[136]{Xe}, which have a low expectation in the ROI, are also included in the flat ER component for this study.
The rate of these ER background components is constrained by a fit to the reconstructed ER energy spectrum in [20, 140]\,\kever outside of the ROI. The best-fit rate is found to be consistent with ancillary measurements of the individual components.
For \sronea, the dominant ER background component originates from the elevated level of \isotope[85]{Kr}, and a subdominant contribution from the K-shell electron capture of \isotope[37]{Ar} ($\SI{\sim2.8}{\kilo\electronvolt}$).
Its rate in \sronea is constrained by extrapolating the \isotope[37]{Ar} decay rate from a reference dataset taken before \sronea.
While backgrounds from \isotope[85]{Kr} and \isotope[37]{Ar} were reduced to a subdominant level in \sroneb, an additional ER background component, with an energy spectrum resembling $\upbeta$ decays of \isotope[3]{H}, was present in both \sronea and \sroneb.
Since this background component only appears in the ROI, its rate was left unconstrained and determined solely from the science data in the ROI, using the \isotope[3]{H} spectral shape.

\tldr{\dec background..}
The LM+LN ($\SI{\sim 6}{\kilo\electronvolt}$) and LL ($\SI{\sim 10}{\kilo\electronvolt}$) shell peaks from the double-electron capture (\dec) of \isotope[124]{Xe} lie within the WIMP search ROI~\cite{XENON:2022evz}.
In \cite{LZ:2025_wimp}, the ER background from \dec was fit with a free \chargeyield parameter to account for a lower \chargeyield due to higher ionization density of electron captures.
In our analysis, however, a likelihood-ratio hypothesis test on \srone data performed after unblinding did not reject the nominal $\upbeta$-yield hypothesis for the LM+LN and LL shell \dec events.
Accordingly, we used the nominal model in which \dec is a part of the flat ER component. 
This strategy was defined before unblinding.
Details on the hypothesis test, as well as the results obtained with the alternative model with free \dec \chargeyield parameters, are provided in the Supplemental Material.

\tldr{NR background, neutrons and neutrinos..}
The NR background mainly originates from radiogenic neutrons produced by spontaneous fission and $(\upalpha, n)$ reactions in detector materials near the LXe target.
A fit to the high-energy $\upgamma$ spectrum suggests that the radioactivity of the inner cryostat flange is significantly higher than expected from material screening results~\cite{XENON:2021radiopurity}.
The neutron expectation from the MC simulations with updated radioactivity is compatible with the data-driven estimate reported in~\cite{XENON:2023wimp}.
While the spatial distribution of the neutron background was derived from the updated MC simulations, the rate was estimated from the neutron sideband, defined by multiple-scatter and single-scatter (SS) events tagged by the NV as in \cite{XENON:2024ap2, XENON:2024neutron_veto}.
The NV tagging efficiency was measured with the same procedure as in \cite{XENON:2024neutron_veto}, resulting in $(55 \pm 2)\%$.
With the validated MC framework, the data-driven constraint, and the updated tagging efficiency, the sideband unblinding yields a neutron background expectation in the WIMP ROI of $0.48 \pm 0.19$ ($0.7 \pm 0.3$) for \sronea (\sroneb) and an updated expectation of $0.7 \pm 0.3$ events for SR0.
Another contribution to the NR background is due to coherent elastic neutrino-nucleus scattering (\cevns) of \isotope[8]{B} solar, atmospheric (atm.), and diffuse supernova neutrino background (DSNB).
Since neutrinos interact weakly with nuclei, they were modeled as SS events, similar to WIMPs.
For all NR components in SR0, the NR response was updated to the best-fit model from the combined SR0+SR1 calibration fit.

\tldr{AC background..}
The AC background was modeled in a data-driven approach as in \cite{XENON:2024cevns,XENON:2023wimp}, using \software{axidence} \cite{axidence}.
The model was validated by the events that satisfy all selection criteria, but fail the GBDT or \stwo width requirements~\cite{XENON:2024ap1}.
The 154 observed events in the sideband were in agreement with the expectation of 137 events.
The uncertainty on the AC background rate in SR1 was calculated as the Poisson uncertainty of the sideband expectation, yielding a relative value of 8.5\,\%.

\tldr{Surface background..}
The surface background in the WIMP ROI originates from $\upbeta$ decay events in the \isotope[210]{Pb} decay chain on the surface of the TPC wall.
These events can lose a significant fraction of ionization electrons, resulting in comparatively smaller \stwo signals.
The surface background model was constructed in a data-driven way as in \cite{XENON:2024ap2}.
The radial modeling was improved by using \isotope[210]{Pb} events with \csone $\in[100, 300]$ PE, which better represent the background than the previously used \isotope[210]{Po} $\upalpha$ events. This update was also applied to SR0.
Events outside the FV were used as a sideband to validate the radial distribution of the surface background model, which demonstrated a good match with the data.

\begin{figure}[]
    \centering
    \includegraphics[width=\columnwidth]{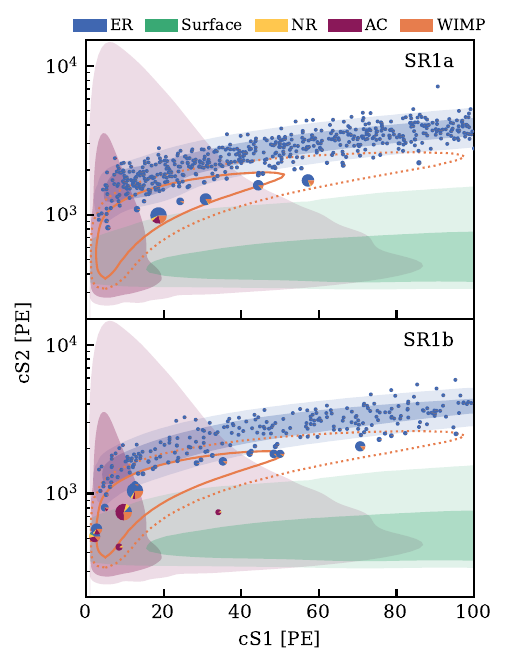}
    \caption{Distribution of events in \csone--\cstwo for \sronea (top) and \sroneb (bottom). All data points are represented as pie charts indicating the fraction of the likelihood from the best-fit model including a 200\,\gevcc WIMP signal, evaluated at the data point. The scatter size is scaled according to the WIMP likelihood fraction for visualization only. In addition, the one (dark shaded, solid line) and two (light shaded, dotted line) $\sigma$ contours of the ER, AC, surface background, and 200\,\gevcc WIMP signal are shown. The NR background follows a distribution similar to that of WIMPs and is therefore not shown separately.}
    \label{fig:cs1_cs2}
\end{figure}

\begin{table*}[ht]
    \centering
    \caption{Expectation values of the nominal (prefit) and best-fit models for SR0, \sronea, and \sroneb (1.09, 0.73, 1.31 tonne-year, respectively), including an unconstrained WIMP signal with a mass of 200\,\gevcc. Connected background colors (c.f. \cref{fig:cs1_cs2}) indicate which components share a scaling parameter, coupling their rates across different science runs. Uncertainties listed in the ``nominal'' columns correspond to the widths of the Gaussian constraints applied in the likelihood.}
    \label{tab:expectation_values}

    \setlength{\tabcolsep}{-0.1pt}
    \renewcommand{\arraystretch}{1.3} 

    \setlength{\arrayrulewidth}{0.8pt}
    \newcommand{\whline}{\arrayrulecolor{white}\hline\arrayrulecolor{black}}

    \newcommand{\rower}{\rowcolor[HTML]{B1C4F2}}
    \newcommand{\rownr}{\rowcolor[HTML]{FFE09E}}
    \newcommand{\rowac}{\rowcolor[HTML]{C6A1B9}}
    \newcommand{\rowsurface}{\rowcolor[HTML]{B8DBC7}}
    \newcommand{\rowwimp}{\rowcolor[HTML]{EAA280}}

    \newcommand{\cellsep}{\cellcolor[HTML]{FFFFFF}} 

    \newcolumntype{W}{>{}m{0.3cm}} 
    \newcolumntype{C}{>{\centering\arraybackslash}m{2.3cm}} 
    \newcolumntype{L}{>{\raggedright\arraybackslash}m{3.3cm}} 

    \begin{tabular}{L  CCW CCW CC}
        \hline
        & \multicolumn{2}{c}{SR0} & & \multicolumn{2}{c}{\sronea} & & \multicolumn{2}{c}{\sroneb}\\
        & Nominal & Best fit & & Nominal & Best fit & & Nominal & Best fit\\
\hline\whline
\rower\cellsep ER (flat) & $ 134 $ & $ 136 \pm 12 $ & \cellsep & $ 430 \pm 30 $ & $ 450 \pm 20 $ & \cellsep & $ 151 \pm 11 $ & $ 154 \pm 10 $\\
\whline
\rower\cellsep ER (\isotope[3]{H}-like) & \cellsep -- & \cellsep -- & \cellsep & $ 62 $ & $ 40 \pm 30 $ & \cellsep & $ 101 $ & $ 80^{+18}_{-17} $\\
\whline
\rower\cellsep ER (\isotope[37]{Ar}) & \cellsep -- & \cellsep -- & \cellsep & $ 58 \pm 6 $ & $ 55 \pm 5 $ & \cellsep & \cellsep -- & \cellsep --\\
\whline\whline
\rownr\cellsep Neutron & \color{black}$ 0.7 \pm 0.3 $ & \color{black}$ 0.6 \pm 0.3 $ & & \color{black}$ 0.47 \pm 0.19 $ & \color{black}$ 0.45 \pm 0.19 $ & & \color{black}$ 0.7 \pm 0.3 $ & \color{black}$ 0.7 \pm 0.3 $\\
\whline
\rownr\cellsep \cevns (solar) & \color{black}$ 0.16 \pm 0.05 $ & \color{black}$ 0.16 \pm 0.05 $ &  & \color{black}$ 0.010 \pm 0.003 $ & \color{black}$ 0.010 \pm 0.003 $ & & \color{black}$ 0.019 \pm 0.006 $ & \color{black}$ 0.019 \pm 0.006 $\\
\whline
\rownr\cellsep \cevns (atm.+DSNB) & \color{black}$ 0.04 \pm 0.02 $ & \color{black}$ 0.04 \pm 0.02 $ &  & \color{black}$ 0.024 \pm 0.012 $ & \color{black}$ 0.024 \pm 0.012 $ & & \color{black}$ 0.05 \pm 0.02 $ & \color{black}$ 0.05 \pm 0.02 $\\
\whline\whline
\rowac\cellsep AC & $ 4.3 \pm 0.9 $ & $ 4.4^{+0.9}_{-0.8} $ & \cellsep & $ 2.12 \pm 0.18 $ & $ 2.10 \pm 0.18 $ & & $ 3.8 \pm 0.3 $ & $ 3.8 \pm 0.3 $\\
\whline\whline
\rowsurface\cellsep Surface & $ 13 \pm 3 $ & $ 11 \pm 2 $ & \cellsep & $ 0.43 \pm 0.05 $ & $ 0.42 \pm 0.05 $ & & $ 0.77 \pm 0.09 $ & $ 0.76 \pm 0.09 $\\
\whline\hline
Total background & $ 152 $ & $ 152 \pm 12 $ & & $ 553 $ & $ 550 \pm 20 $ & & $ 257 $ & $ 239 \pm 15 $ \\
\whline\whline
\rowwimp\cellsep WIMP (200\,\gevcc) & -- & $ 1.8 $ & & -- & $ 1.1 $ & & -- & $ 2.1 $\\
\whline\hline
Observed & \multicolumn{2}{c}{152} & & \multicolumn{2}{c}{560} & & \multicolumn{2}{c}{245} \\
\whline\hline
    \end{tabular}
\end{table*}

\textit{Statistical inference\textemdash}For the statistical analysis of the dataset, we used a log-likelihood-based test statistic with the distributions obtained via \toymc simulations, as recommended in~\cite{Baxter:2021pqo} and detailed in~\cite{XENON:2024ap2}.
The computations were performed with the \software{alea} framework~\cite{alea}.
The likelihood function $\mathcal{L}(\sigma, \boldsymbol{\theta})$ depends on the WIMP-nucleon cross section $\sigma\geq 0$, which is the parameter of interest, and a set of nuisance parameters $\boldsymbol{\theta}$. It factorizes into three components: $\mathcal{L}(\sigma, \boldsymbol{\theta}) = \mathcal{L}_\mathrm{sci}(\sigma, \boldsymbol{\theta}) \times \mathcal{L}_\mathrm{cal}(\boldsymbol{\theta})\times \mathcal{L}_\mathrm{anc}(\boldsymbol{\theta}).$
The science search likelihood function $\mathcal{L}_\mathrm{sci}$ itself factorizes into six parts, corresponding to SR0, \sronea, and \sroneb, each subdivided into near- and far-wire regions.
All six are extended unbinned likelihood functions, which model the data in \csone--\cstwo--\R for the far-wire region and in \csone--\cstwo for the near-wire region.
$\mathcal{L}_\mathrm{cal}$ consists of two unbinned likelihood functions in \csone--\cstwo modeling the ER calibration datasets in SR0 and SR1, and $\mathcal{L}_\mathrm{anc}(\boldsymbol{\theta})$ is a product of Gaussian constraints for nuisance parameters from ancillary measurements.
The background and signal components are listed in \cref{tab:expectation_values}.
Apart from the background expectation values, the set of nuisance parameters comprises the WIMP-mass-dependent relative signal efficiency and four ER shape parameters (two for each SR) that modify the shape in \csone--\cstwo of the different ER components. 
These parameters are tightly constrained by $\mathcal{L}_\mathrm{cal}$.
The relative signal rate uncertainty is 15\,\% (6\,\%, 4\,\%) in SR0 (\sronea, \sroneb) for WIMP masses above $\sim$100\,\gevcc and becomes larger for smaller masses.
The rate uncertainty in SR1 is smaller than in SR0 due to a smaller selection efficiency uncertainty.

We employed power-constrained limits (PCL)~\cite{Cowan:2011an,Baxter:2021pqo} to prevent excluding regions of parameter space where our sensitivity is low, which could otherwise occur due to statistical fluctuations or systematic effects.
In~\cite{XENON:2023wimp}, a conservative power threshold of 0.5 was chosen after identifying an error in the definition of power in~\cite{Baxter:2021pqo}, effectively truncating the upper limits at the median of the sensitivity band.
We have investigated the PCL behavior with toy data, specifically in scenarios involving a shift in the ER event distribution, increased background rates, or increased background uncertainties.
These studies revealed no issues that would disqualify a lower power threshold of 0.16.
The corresponding truncation of the limits at the $-1\sigma$ quantile of the sensitivity band allows for a direct comparison with other experiments~\cite{PandaX:2025_wimp_search,LZ:2025_wimp}.

The SR1 signal region unblinding was performed in two steps.
First, events in a small region above the median of the NR event distribution with energies above $5\,\kever$ were unblinded, containing about 7.5\,\% of expected events from a $1\,\tevcc$ WIMP signal.
This initial step allowed us to investigate potential excessive downward leakage of ER events, as previously observed in~\cite{XENON:2023wimp}.
The results of the first unblinding step showed no discrepancy with the nominal model.
In the second step, all data in the ROI were unblinded.
The regions in cS1--cS2 are indicated in \cref{fig:pie_plot_dec} in the Supplemental Material.

\begin{figure}[]
    \centering
    \includegraphics[width=\columnwidth]{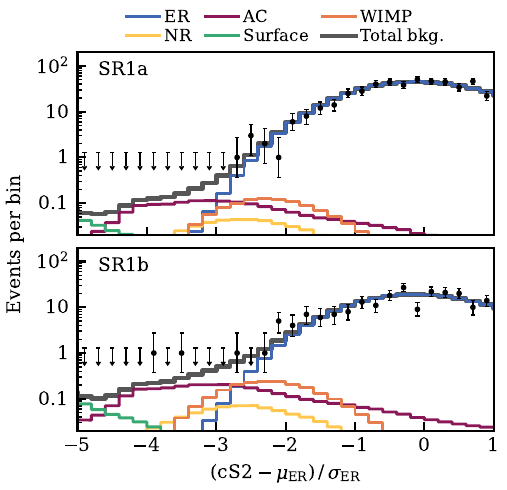}
    \caption{Distribution in cS2 of the observed data and the best-fit model including an unconstrained 200\,\gevcc SI WIMP component in SR1a (top) and SR1b (bottom). The cS2 is normalized by subtracting the median $\mu_{\mathrm{ER}}$ and dividing by the standard deviation $\sigma_{\mathrm{ER}}$ of the ER distribution along cS1. The gray histogram represents the total background expectation. Black dots represent observed event counts, while triangles mark bins with zero events, both with Poisson confidence intervals.}
    \label{fig:cs2_projection}
\end{figure}

\textit{Results\textemdash}After unblinding, we observed 560 events in \sronea and 245 in \sroneb within the ROI, of which 14 and 13, respectively, lie in the previously blinded region (with two additional events in \sroneb, which were already unblinded in \cite{XENON:2024cevns}).
The distribution of all events in \csone--\cstwo is shown in \cref{fig:cs1_cs2}, the corresponding plot for SR0 is Figure 3 in~\cite{XENON:2023wimp}.
\cref{tab:expectation_values} shows the best-fit expectation values for all SRs.
We performed independent GOF tests on \sronea and \sroneb data.
The tests were defined before unblinding with p-value thresholds of 2.5\,\% to reject the best-fit model.
An unbinned Anderson-Darling (AD) test \cite{anderson1952asymptotic} was performed in the \cstwo dimension, after subtracting the \csone-dependent median of the best-fit model.
The data and best-fit models in this space are visualized in \cref{fig:cs2_projection}.
Additionally, we performed binned Poisson $\chi^2$ tests using an equiprobable binning scheme in \csone--\cstwo.
We found no indication of mismodeling in any of the tests, with p-values of 0.34 (\sronea) and 0.85 (\sroneb) for the AD test, and 0.33 (same for \sronea and \sroneb) for the Poisson $\chi^2$ test.
We also performed an \xy-plane spatial uniformity test of the unblinded events in SR1.
It is quantified by the fraction of events in the densest quarter and densest half of the \xy plane.
We found no indication of a spatial asymmetry for SR1.

\begin{figure}[]
    \centering
    \includegraphics[width=\columnwidth]{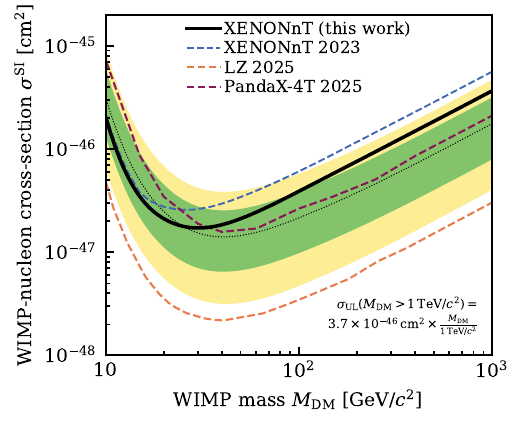}
    \caption{Upper limits on the SI WIMP-nucleon cross section (90\,\% CL) as a function of the WIMP mass (black line). The sensitivity band is indicated by the region containing 68\,\% (green shaded) and 95\,\% (yellow shaded) of expected upper limits under the background-only hypothesis as well as their median (dotted line). In addition, we show published results from XENONnT using only SR0 data~\cite{XENON:2023wimp}, LZ~\cite{LZ:2025_wimp}, and PandaX-4T~\cite{PandaX:2025_wimp_search}. For all, a PCL with a power threshold of 0.16 is used (XENONnT SR0 limit recast accordingly).}
    \label{fig:limit_si}
\end{figure}

\tldr{WIMP discovery p-value and limits}
The local WIMP discovery significance was evaluated for WIMP masses between 10\,\gevcc and 1\,\tevcc.
We found no significant excess above backgrounds with the lowest p-value of 0.13 for a WIMP mass of 1\,\tevcc.
We thus report the WIMP-mass-dependent upper limits of the SI WIMP-nucleon cross section at 90\,\% confidence level (CL), shown in \cref{fig:limit_si} together with the sensitivity band.
As we observed no limit below the $-1\,\sigma$ sensitivity band, no adjustment is needed to satisfy the power constraint requirement.
The most stringent limit on the cross section is $1.7\,\times\,10^{-47}\,\mathrm{cm^2}$ for a WIMP mass of 30\,\gevcc.
For WIMP masses above $\sim200$\,\gevcc the limit scales like $M_{\mathrm{WIMP}}/(1\,\tevcc) \,\times\, 3.7\,\times\,10^{-46}\,\mathrm{cm^2}$.
The Supplemental Material includes the limit expressed in terms of the number of WIMP events and the SR1-only result, as well as the limits for spin-dependent (SD) WIMP-nucleon coupling.


\textit{Summary and outlook\textemdash}In summary, we have performed a blind analysis of a combined SR0+SR1 dataset from XENONnT with a total of 186.5 (95.1) days of live time in SR1 (SR0), resulting in a total exposure of 3.1 tonne-year.
We found no significant excess above background and placed new limits on the SI WIMP-nucleon interaction cross section, with an improvement of approximately a factor of 1.5 for WIMP masses above 30\,\gevcc compared to the SR0-only results.
Running the radon distillation system at its full capacity, we have significantly reduced our \isotope[222]{Rn} concentration by a factor 1.9 (1.7) in SR1a (SR1b), resulting in a record-low ER background from \isotope[214]{Pb} $\upbeta$ decays.

The experiment continues to take data, with an increased NV tagging efficiency due to 0.05\,\% by-weight gadolinium loading, giving an expected factor $\sim$~2 reduction of the neutron background.
The recent installation of a charcoal purifier shows promise in removing photoionizing impurities, which is expected to result in a lower AC background rate.

\textit{Acknowledgments\textemdash}We gratefully acknowledge support from the National Science Foundation, Swiss National Science Foundation, German Ministry for Education and Research, Max Planck Gesellschaft, Deutsche Forschungsgemeinschaft, Helmholtz Association, Dutch Research Council (NWO), Fundacao para a Ciencia e Tecnologia, Weizmann Institute of Science, Binational Science Foundation, Région des Pays de la Loire, Knut and Alice Wallenberg Foundation, Kavli Foundation, JSPS Kakenhi, JST FOREST Program, and ERAN in Japan, Tsinghua University Initiative Scientific Research Program, DIM-ACAV+ Région Ile-de-France, and Istituto Nazionale di Fisica Nucleare. This project has received funding/support from the European Union’s Horizon 2020 research and innovation program under the Marie Skłodowska-Curie grant agreement No 860881-HIDDeN.
We gratefully acknowledge support for providing computing and data-processing resources of the Open Science Pool and the European Grid Initiative, at the following computing centers: the CNRS/IN2P3 (Lyon - France), the Dutch national e-infrastructure with the support of SURF Cooperative, the Nikhef Data-Processing Facility (Amsterdam - Netherlands), the INFN-CNAF (Bologna - Italy), the San Diego Supercomputer Center (San Diego - USA) and the Enrico Fermi Institute (Chicago - USA). We acknowledge the support of the Research Computing Center (RCC) at The University of Chicago for providing computing resources for data analysis.
We thank the INFN Laboratori Nazionali del Gran Sasso for hosting and supporting the XENON project.

\textit{Data availability\textemdash}The data that support the findings of this article are openly available~\cite{data_release}.

\bibliography{reference}

\clearpage



\appendix
\section{Supplemental Material}
\setcounter{equation}{0}
\setcounter{page}{1}
\makeatletter
\renewcommand{\theequation}{S\arabic{equation}}
\renewcommand{\thefigure}{S\the\numexpr\value{figure}-4\relax} 
\renewcommand{\thetable}{S\the\numexpr\value{table}-1\relax}

\subsection{Charge yield of \isotope[124]{Xe} double-electron capture events}
\label{sec:dec}
The \dec of \isotope[124]{Xe} contributes to the ER background in the WIMP ROI with an expected number of ($4.5\pm 0.7$)\,events in SR0 and ($9.1\pm1.4$)\,events in SR1 from the LM + LN ($\SI{\sim 6}{\kilo\electronvolt}$) and LL ($\SI{\sim 10}{\kilo\electronvolt}$) peaks.
The event rate was constrained via the KK and KL-shell capture peaks outside the ROI, using the branching fractions 72.4\,\% (KK), 20.0\,\% (KL), 1.4\,\% (LL), and 0.7\,\% (LM + LN)~\cite{XENON:2022evz}.
The de-excitation cascade following the capture can lead to an increased ionization density, which may result in a higher recombination probability and thus lower \chargeyield compared to $\upbeta$ decay events.
A charge suppression factor of $\sim0.9$ was measured for L-shell electron capture in \isotope[127]{Xe} ($\SI{\sim 5.2}{\kilo\electronvolt}$) at drift fields above 250\,V/cm~\cite{xelda_ec_2021},
but no such measurement exists for \dec at the 23\,V/cm field used in this search.

Introducing reduced charge yields as nuisance parameters could bias upper limits by absorbing leakage from other ER backgrounds, artificially lowering them.
Thus, we first tested the hypothesis of a reduced \chargeyield for the \dec peaks at 23\,V/cm using SR1 science data.
For this, an alternative likelihood function was defined, modeling the \dec peaks as separate components with $r_{\mathrm{LL}} = \mathrm{\chargeyield_{LL}}/\mathrm{\chargeyield}_\upbeta$ and $r_{\mathrm{LM}} = \mathrm{\chargeyield_{LM}}/\mathrm{\chargeyield}_\upbeta$ as free-floating parameters, without ancillary constraints.

\begin{figure}[h]
    \centering
    \includegraphics[width=\linewidth]{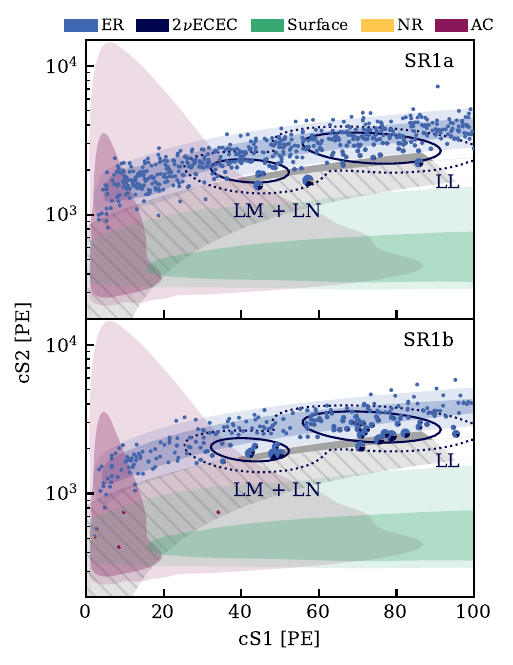}
    \caption{The cS1--cS2 distribution for \srone, from \Cref{fig:cs1_cs2}. The model was modified to include the LM+LN and LL \dec peaks with unconstrained \chargeyield while excluding the WIMP component. The regions corresponding to the first (solid gray) and second (hatched gray) steps of unblinding are also indicated.}
    \label{fig:pie_plot_dec}
\end{figure}

A likelihood-ratio test compared the nominal \chargeyield hypothesis ($r_{\mathrm{LL}} = r_{\mathrm{LM}} = 1$) against an unconstrained alternative.
A test size of $\upalpha = 5\,\%$ was chosen to limit the rate of false WIMP discovery for several scenarios with reduced \dec charge yields when not modeled appropriately ($>3\,\sigma$ false WIMP discovery rate below 0.3\,\%).
After unblinding the data, a p-value of 0.09 was observed, so the nominal \chargeyield hypothesis was not rejected.
As a result, the nominal model without charge yield suppression, as described in the main text, was used for the WIMP search.
The best-fit model using only SR1 science data is illustrated in \cref{fig:pie_plot_dec}, yielding $r_{\mathrm{LL}}\approx0.9$ and $r_{\mathrm{LM}}\approx0.8$.
The best-fit results using the combined SR0+SR1 dataset are $r_{\mathrm{LL}}=0.80^{+0.08}_{-0.04}$ and $r_{\mathrm{LM}}=0.72^{+0.11}_{-0.04}$.

\begin{figure}[h!]
    \centering
    \includegraphics[width=\columnwidth]{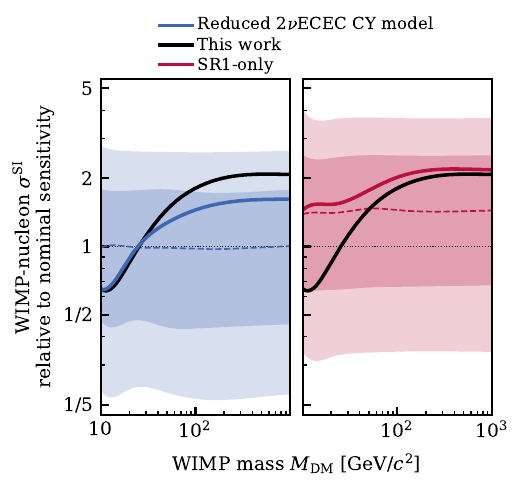}
    \caption{Sensitivity and upper limits for the model incorporating reduced \dec \chargeyield parameters (left) and SR1-only (right), shown relative to the sensitivity of the nominal model for the full SR0+SR1 dataset. The dark (light) shaded region contains $68\%$ ($95\%$) of expected upper limits under the background-only hypothesis and the dashed colored lines indicate respective medians.
    The solid colored lines represent the observed upper limits for each model. The black lines correspond to the observed upper limit of the nominal model shown in \cref{fig:limit_si}.}
    \label{fig:limit_comparison}
    \smallskip
    \includegraphics[width=\columnwidth]{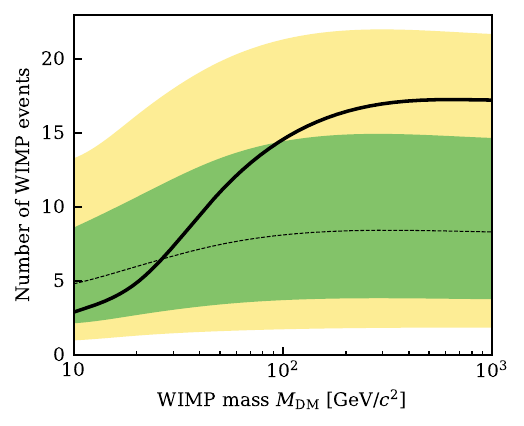}
    \caption{Sensitivity band (yellow-green), median upper limit (dashed line), and observed upper limit (solid line) for the SI SR0+SR1 WIMP search, all expressed in terms of the number of WIMP events.}
    \label{fig:n_ev_limit}
\end{figure}

For comparison, we also report the upper limits using an alternative statistical model with yields as nuisance parameters.
The LM+LN peak yield was loosely constrained to $r_{\mathrm{LM}}=0.9\pm0.1$, based on a conservative extrapolation~\cite{xelda_ec_2021}.
The LL peak was left unconstrained.
The resulting upper limits are shown in \cref{fig:limit_comparison} (left), plotted relative to the sensitivity of the nominal model.
At high WIMP masses, the upper limit is a factor of 1.3 more stringent compared to the nominal model.
The sensitivity was computed with a nominal value of $r_{\mathrm{LL}}=0.8$.
The competing effects of increased background leakage and the added degrees of freedom largely cancel, leaving the sensitivity mostly unchanged.

\subsection{SR1-only and WIMP event count limits}

Alongside the upper limits derived from the combined SR0+SR1 dataset, we also present the results obtained using only the SR1 dataset.
These are shown in \cref{fig:limit_comparison} (right) relative to the combined result.
For all masses, the upper limit is above the median, indicating a slight overfluctuation.

Additionally, we present the sensitivity and limits expressed in terms of the corresponding number of WIMP events in \cref{fig:n_ev_limit}.

\subsection{Results on spin-dependent WIMP-nucleon interactions}
\begin{figure}[H]
    \centering
    \includegraphics[width=\columnwidth]{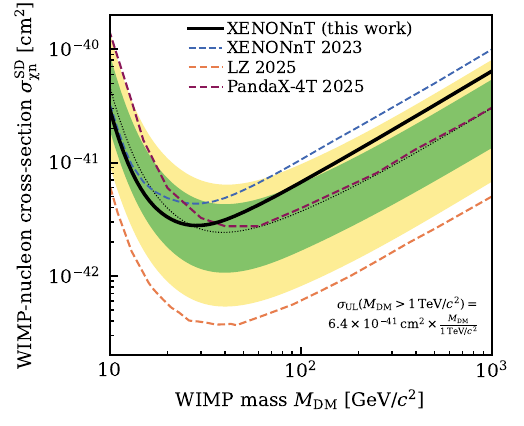}
    \includegraphics[width=\columnwidth]{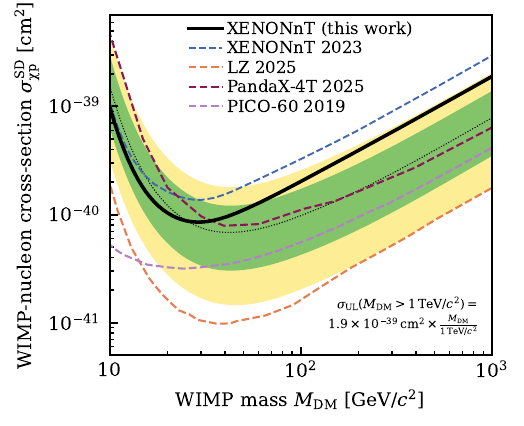}
   \caption{Upper limits on the spin-dependent WIMP-nucleon cross section, for the ``neutron-only'' (top) and ``proton-only'' (bottom) cases. As in \cref{fig:limit_si}, we show published results from XENONnT using only SR0 data~\cite{XENON:2023wimp} (limit recast with a power threshold of 0.16), LZ~\cite{LZ:2025_wimp}, and PandaX-4T~\cite{PandaX:2025_wimp_search}. Results from PICO-60~\cite{PICO:2019vsc} are shown in the bottom plot only.}
   \label{fig:limit_sd}
\end{figure}

Using the combined dataset in this analysis, we also computed the upper limit on cross sections for spin-dependent (SD) WIMP-nucleon interactions. Since the interaction vanishes for zero nuclear spin, only two xenon isotopes, \isotope[129]{Xe} and \isotope[131]{Xe}, contribute. The nuclear structure factors used in this analysis are the medians from~\cite{Klos:2013rwa}. A more recent study~\cite{Hoferichter:2020osn} provides improved results for the SD structure factors which will be used in future analyses. We report limits with respect to the astrophysical parameters recommended in~\cite{Baxter:2021pqo}. The resulting limits are shown in \cref{fig:limit_sd}.

\end{document}